\documentclass[superscriptaddress,
twocolumn,showpacs,preprintnumbers,amsmath,amssymb]{revtex4}
\usepackage{graphicx}% Include Figure files
\usepackage{dcolumn}% Align table columns on decimal point
\usepackage{bm}% bold math
\usepackage{latexsym}
\begin{document}
%%%%%%%%%%%%%%%%%%%%%%%%%%%%%%%%%%%%%%%%%%%%%%%%%%%%%%%%%%%%%%%%
\title{Transcriptional bursts: a unified model of machines and mechanisms}
%\shorttitle{Transcriptional burst statistics}
\author{Tripti Tripathi}
\affiliation{Department of Physics, Indian Institute of Technology,
Kanpur 208016, India.}
\author{Debashish Chowdhury}
\affiliation{Department of Physics, Indian Institute of Technology,
Kanpur 208016, India.}
%\date{\today}%
%%%%%%%%%%%%%%%%%%%%%%%%%%
\begin{abstract}
{\it Transcription} is the process whereby RNA molecules are polymerized 
by molecular machines, called RNA polymerase (RNAP), using the 
corresponding DNA as the template. Recent {\it in-vivo} experiments 
with single cells have established that transcription takes place in 
``bursts'' or ``pulses''. In this letter we present a model that captures 
not only the mechano-chemistry of individual RNAPs and their steric 
interactions but also the switching of the gene between the ON and OFF 
states. This model accounts for the statistical properties of the 
transcriptional bursts. It also shows how the quantitative features of 
the distributions of these bursts can be tuned by controlling the 
appropriate steps of operation of the RNAP machines. 
\end{abstract}
%%%%%%%%%%%%%%%%%%%%%%%%%%
\maketitle
%%%%%%%%%%%%%%%%%%%%%%%%%%%%%%%%%%%%%%%%%%%%%%%%%%%%%%%%%%%%

Genetic messages are encoded chemically in DNA. During gene expression 
this message is first {\it transcribed} into mRNA and then, from it, 
{\it translated} into proteins by well coordinated operation of  
intracellular machineries \cite{alberts}. The two machines, which play 
key roles in transcription and translation are the RNA polymerase (RNAP)
\cite{wangrev} and the ribosome \cite{spirinbook,spirin02}, respectively. 
Each of these machines is like a mobile workshop that  synthesizes a 
bio-polymer according to a template which also serves as the track 
for the movement of the workshop. Because of the probabilistic nature 
of the steps of gene expression, the number of mRNA and protein 
molecules corresponding to a single gene fluctuate randomly (see 
ref.\cite{raser05,kaern05,paulsson05,maheshri07,oudenaarden07} for reviews).
It has been observed experimentally \cite{golding05,chubb06,raj06,xie08} 
that relatively long periods $T_{off}$ of transcriptional inactivity 
are interspersed with brief periods $T_{on}$ of transcriptional 
``bursts''. Several statistical properties of these random ``bursts'' 
(or, ``pulses'') have been used to characterize the temporal pattern 
in transcriptional events \cite{golding05,chubb06,raj06}. 

%%%%%%%%%%%%%%%%%%%%%%%%%%%%%%%%%%%%%%%%%%%%%%%%%%%%%%%%%%
\begin{figure}[t]
\begin{center}
\includegraphics[angle=-90,width=0.9\columnwidth]{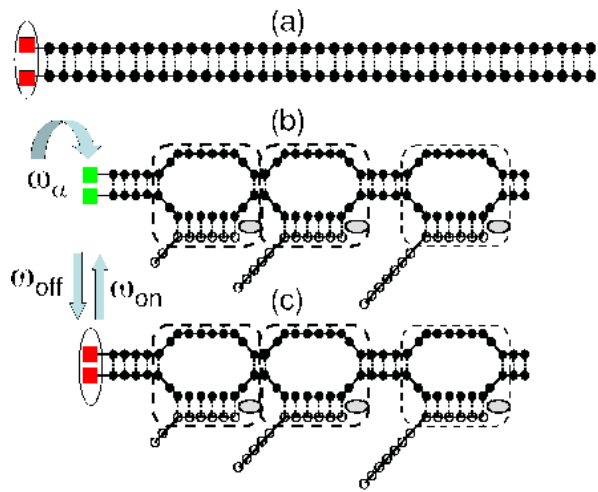}
\end{center}
\caption{(Color online) A pictoral depiction of the model. The three 
dashed squares represent three TECs. The solid lines connecting filled 
circles represent the single strands of DNA while the string of open 
circles denotes the elongating RNA molecule. The dashed lines connecting 
the circles denote the unbroken non-covalent bonds between the complementary 
subunits on the DNA and RNA strands. Each of the grey ovals represents 
the catalytic site on the corresponding RNAP. The green and red squares 
indicate the ON and OFF states of the gene. The rates of the transitions 
between the ON and the OFF states as well as the rate of transcription 
initiation in the ON state of the gene are also shown explicitly.
}
\label{fig-model}
\end{figure}
%%%%%%%%%%%%%%%%%%%%%%%%%%%%%%%%%%%%%%%%%%%%%%%%%%%%%%%%%%

%%%%%%%%%%%%%%%%%%%%%%%%%%%%%%%%%%%%%%%%%%%%%%%%%%%%%%%%%%
\begin{figure}[t]
\begin{center}
\includegraphics[angle=-90,width=0.9\columnwidth]{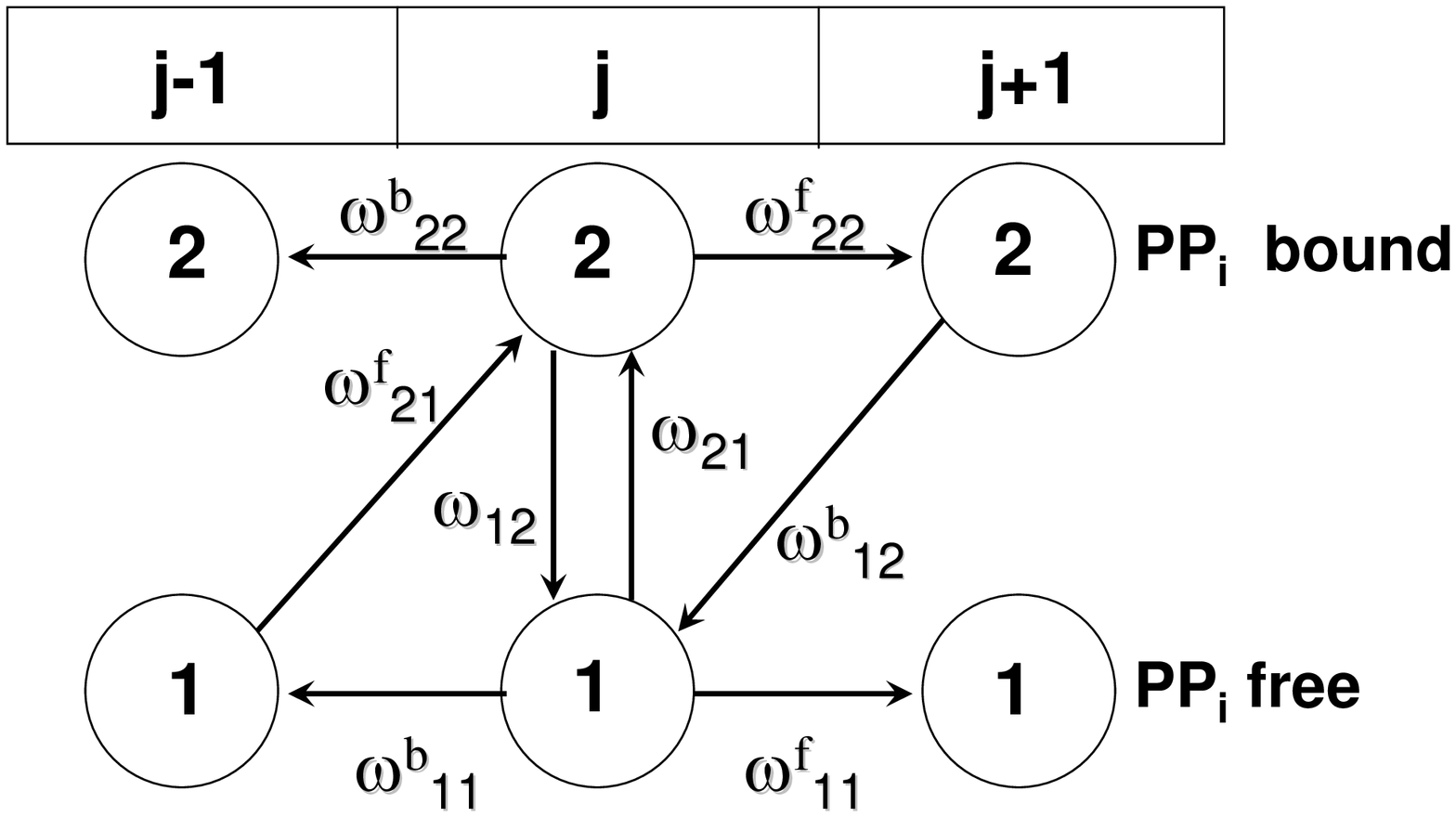}
\includegraphics[angle=-90,width=0.9\columnwidth]{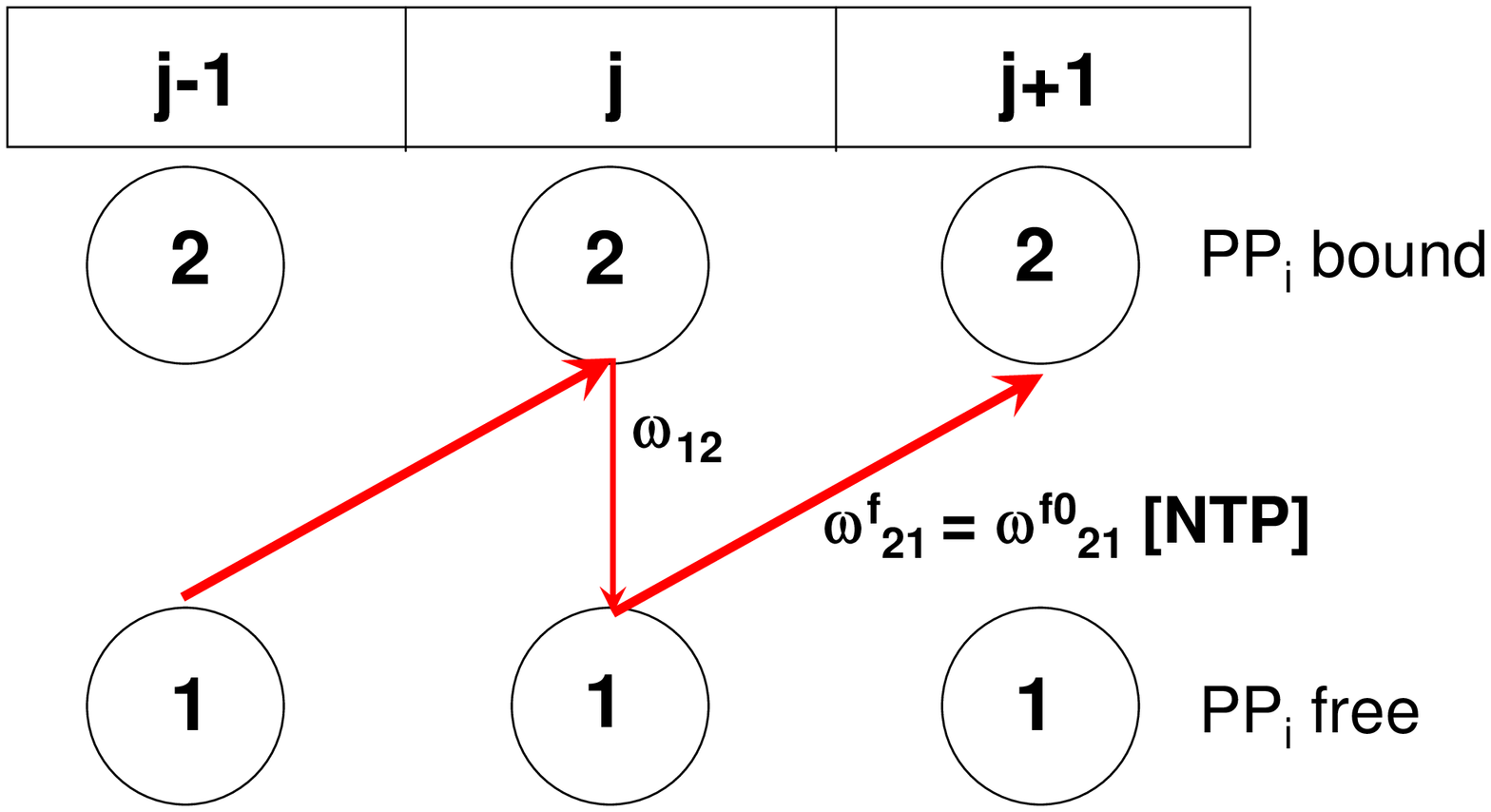}
\end{center}
\caption{(Color online) Allowed mechano-chemical transitions of individual 
RNAPs during the elongation stage in our model. The indices $j-1, j, j+1$ 
denote an arbitrary sequence of three nucleotides on the template DNA. 
The encircled symbols $1$ and $2$ denote the two possible chemical 
states; no pyrophosphate (PP$_{i}$) is bound to the RNAP in the state 
$1$ whereas PP$_{i}$-bound state is labelled by the index $2$.
The directions of the arrows and the associated symbols indicate the 
possible transitions and the corresponding rate constants, respectively. 
Elongation of the nascent RNA transcript is accompanied by forward 
movement of the RNAP whereas backward movements of the RNAP correspond 
to depolymerization of the RNA. The full model, shown in (a), allows 
mechano-chemical transitions which branch off the dominant pathway of 
an individual RNAP shown in (b).  
}
\label{fig-mechchem1}
\end{figure}
%%%%%%%%%%%%%%%%%%%%%%%%%%%%%%%%%%%%%%%%%%%%%%%%%%%%%%%%%%

%%%%%%%%%%%%%%%%%%%%%%%%%%%%%%%%%%%%%%%%%%%%%%%%%%%%%%%%%%
%\begin{figure}[t]
%\begin{center}
%\includegraphics[angle=-90,width=0.9\columnwidth]{tcrevfig8.eps}
%\end{center}
%\caption{(Color online) The dominant pathway of the RNAP in the 
%mechano-chemical cycle of an individual RNAP in our model.
%}
%\label{fig-mechchem2}
%\end{figure}
%%%%%%%%%%%%%%%%%%%%%%%%%%%%%%%%%%%%%%%%%%%%%%%%%%%%%%%%%%

Qualitatively similar bursts of transcriptional activities 
have been observed in both prokaryotes and eukaryotes. 
Some possible mechanisms of transcriptional burst have been 
suggested. Transcription, i.e., the process of synthesis of RNA from 
the corresponding DNA template, can be broadly divided into three 
stages, namely, {\it initiation}, {\it elongation} and {\it termination}.
When the gene is switched ``ON'', initiation of transcription by RNAPs 
can take place till the gene  switches back to the ``OFF'' state 
\cite{golding06}. 
Unbinding and binding of transcription repressor molecules can 
give rise to such switching ``ON'' and ``OFF'' of a bacterial gene. 
In eukaryotic cells, chromatin remodelling enzymes can act as activators 
of transcription. Even if the gene does not switch OFF, burst-like 
transcriptional activities are possible if several RNAPs queue up 
behind a stalled RNAP and then, suddenly, the stalled RNAP gets 
reactivated \cite{artsimovitch00,shundrovsky04}. The latter becomes 
very pronounced \cite{liverpool} when pausing is caused by backtracking 
of the RNAP \cite{galburt07}.

%%%%%%%%%%%%%%%%%%%%%%%%%%%%%%%%%%%%%%%%%%%%%%%%%%%%%%%%%%
\begin{figure}[t]
\begin{center}
\includegraphics[width=0.9\columnwidth]{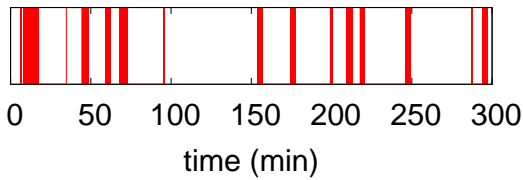}
\end{center}
\caption{(Color online) A typical time series of the transcriptional 
events in the model; each vertical bar corresponds to the completion 
of polymerization of a RNA molecule.
}
\label{fig-timeseries}
\end{figure}
%%%%%%%%%%%%%%%%%%%%%%%%%%%%%%%%%%%%%%%%%%%%%%%%%%%%%%%%%%

%%%%%%%%%%%%%%%%%%%%%%%%%%%%%%%%%%%%%%%%%%%%%%%%%%%%%%%%%%
\begin{figure}[t]
\begin{center}
\includegraphics[width=0.9\columnwidth]{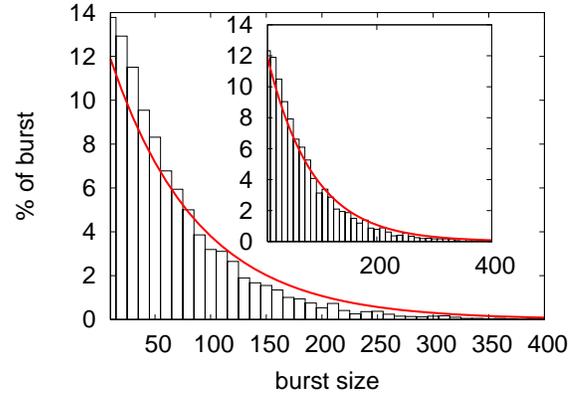}
\end{center}
\caption{(Color online) The distribution of the sizes of the 
transcriptional bursts in our model plotted using $\Delta t = 0.5$ min 
($\Delta t = 2.5$ min in the inset). The continuous line (red) is  
obtained from the theoretically predicted form (\ref{eq-sizeburst}).  
The data points, plotted as bars, were obtained from computer simulations 
of the model. 
}
\label{fig-sizeburst}
\end{figure}
%%%%%%%%%%%%%%%%%%%%%%%%%%%%%%%%%%%%%%%%%%%%%%%%%%%%%%%%%%

%%%%%%%%%%%%%%%%%%%%%%%%%%%%%%%%%%%%%%%%%%%%%%%%%%%%%%%%%%
\begin{figure}[t]
\begin{center}
\includegraphics[width=0.9\columnwidth]{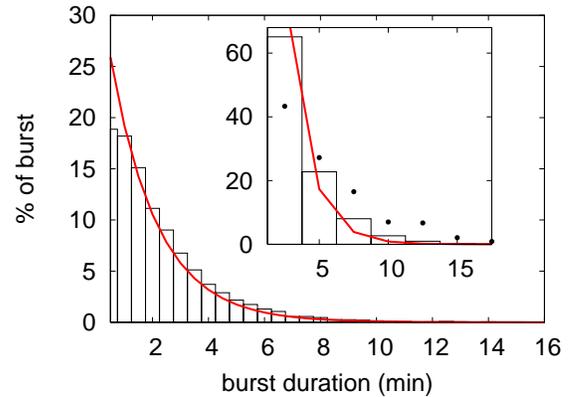}
\end{center}
\caption{(Color online) Distribution of the durations of transcriptional 
bursts in our model plotted using $\Delta t = 0.5$ min ($\Delta t = 2.5$ 
min in the inset). The continuous lines (red) are
obtained from the theoretically predicted form (\ref{eq-durburst}).
The data points, plotted as bars, were obtained from computer simulations 
of the model. The black dots in the inset represent the experimental 
data reported by Chubb et al. in ref.\cite{chubb06}.
}
\label{fig-durburst}
\end{figure}
%%%%%%%%%%%%%%%%%%%%%%%%%%%%%%%%%%%%%%%%%%%%%%%%%%%%%%%%%%

To our knowledge, the switching of the gene between the active (ON) 
and inactive (OFF) states is a common feature of almost all the models 
of transcriptional noise \cite{kepler01,lipniacki06}. 
But, these models capture the entire processes of RNA production 
by a single effective rate constant. In contrast, models developed 
to understand the operational mechanisms of RNAP motors 
\cite{julicherrnap98,osterrnap98,sousa06,mdwang07,nudler05,tadigotla,woo06,liverpool,tripathi08} 
explicitly describe the different stages of transcription, namely, 
initiation, elongation and termination, but (except for 
ref.\cite{tripathi08}) do not address the question of temporal 
fluctuations in transcription. The main aim of this letter is to 
combine the key features of these two types of models within a single 
unified theoretical framework.

%%%%%%%%%%%%%%%%%%%%%%%%%%%%%%%%%%%%%%%%%%%%%%%%%%%%%%%%%%
\begin{figure}[t]
\begin{center}
\includegraphics[width=0.9\columnwidth]{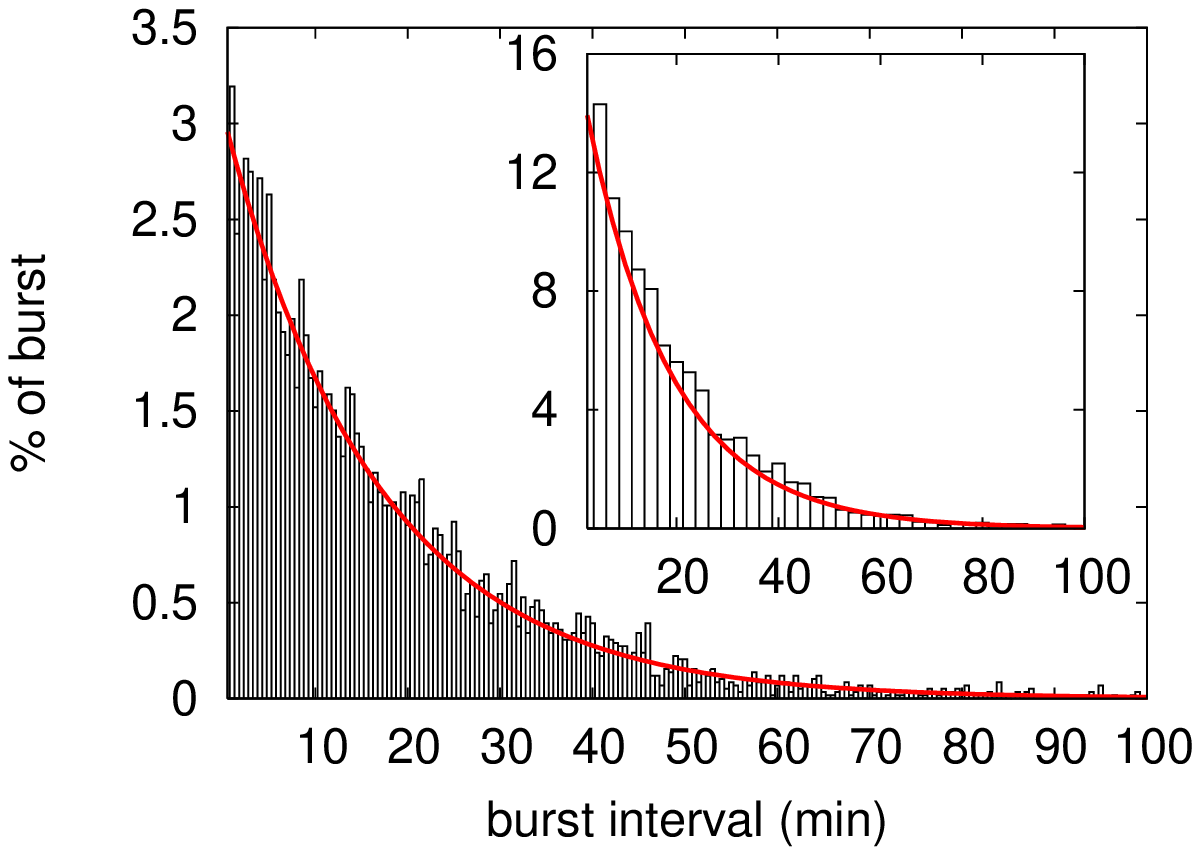}
\end{center}
\caption{(Color online) Distribution of the intervals between successive 
bursts of transcriptional activities in our model plotted using 
$\Delta t = 0.5$ min ($\Delta t = 2.5$ min in the inset). 
The continuous lines (red) are obtained from the theoretically predicted 
form (\ref{eq-intburst}). The data points, plotted as bars, were obtained 
from computer simulations of the model.
}
\label{fig-intburst}
\end{figure}
%%%%%%%%%%%%%%%%%%%%%%%%%%%%%%%%%%%%%%%%%%%%%%%%%%%%%%%%%%

%%%%%%%%%%%%%%%%%%%%%%%%%%%%%%%%%%%%%%%%%%%%%%%%%%%%%%%%%%
\begin{figure}[t]
\begin{center}
\includegraphics[width=0.9\columnwidth]{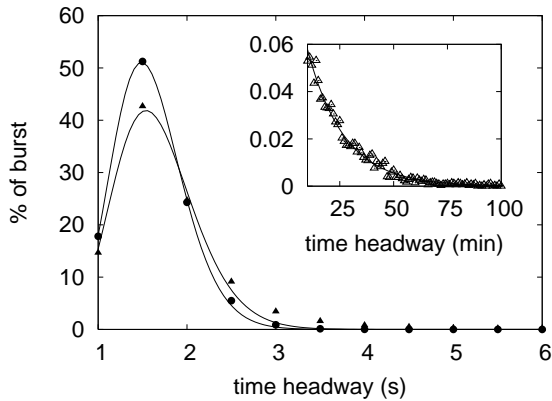}
\end{center}
\caption{Distribution of the time headways between the successive RNAPs 
in our model (triangles). The filled circles show the time heaway 
distribution in the hypothetical scenario (which was assumed in 
ref.\cite{tripathi08}) where the gene remains in the ON state throughout 
the period of computation. The inset shows the long tail of the time 
headway distribution which corresponds to the time gaps between the 
successive bursts.
}
\label{fig-thrnap}
\end{figure}
%%%%%%%%%%%%%%%%%%%%%%%%%%%%%%%%%%%%%%%%%%%%%%%%%%%%%%%%%%

%%%%%%%%%%%%%%%%%%%%%%%%%%%%%%%%%%%%%%%%%%%%%%%%%%%%%%%%%%
\begin{figure}[t]
\begin{center}
\includegraphics[width=0.9\columnwidth]{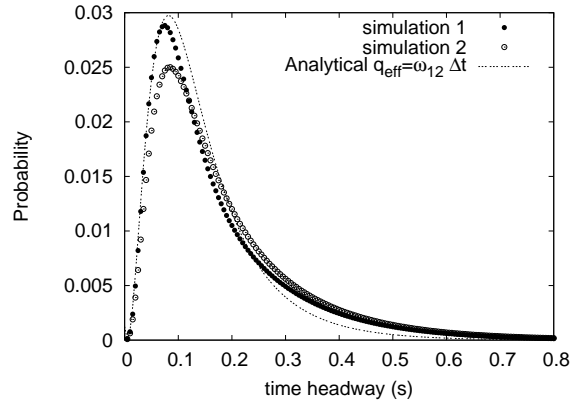}
\end{center}
\caption{Comparison between the approximate analytical expression 
(\ref{eq-thtasep}) for the TH distribution (denoted by the line) 
and the corresponding simulation data (denoted by the discrete 
data points) in the special case where $r = 1$; periodic boundary 
conditions are imposed and the gene remains ``ON'' during the 
entire duration of observation. For the two sets of simulation 
data labelled as ``simulation 1'' and ``simulation 2'' the 
mechano-chemical transitions shown in figs.\ref{fig-mechchem1}(b) 
%mechano-chemical transitions shown in figs.\ref{fig-mechchem2} 
and \ref{fig-mechchem1}(a), respectively, have been used. 
}
\label{fig-anacomp}
\end{figure}
%%%%%%%%%%%%%%%%%%%%%%%%%%%%%%%%%%%%%%%%%%%%%%%%%%%%%%%%%%

More specifically, we extend our recent model of RNAP traffic 
\cite{tripathi08} by allowing the gene to switch between the ``ON'' 
and ``OFF'' states. In other words, this extended model explicitly 
describes the following processes:
(i) switching ``ON'' and ``OFF'' of the gene, 
(ii) initiation, elongation and termination of transcription, 
(iii) mechano-chemical cycles of the individual RNAP motors in the 
elongation stage, and 
(iv) congestion of traffic of RNAPs caused by their steric interaction. 
Consequently, this model can predict the contributions of the processes 
(i)-(iv) on transcriptional noise; estimation of the contributions 
made by the processes (ii)-(iv) was beyond the scope of all the earlier 
models of transcriptional noise.

Carrying out computer simulations of this model we obtain the time series 
of the transcriptional events. 
We sort the transcriptional events of each time series obtained from our 
simulations into ``bursts'' by using well-defined criteria (which we 
describe below). We compare various statistical properties of these  
theoretically predicted transcriptional bursts with the corresponding 
experimental results. We then suggest an alternative statistical analysis 
of the transcriptional noise in terms of some new distributions which are 
motivated by superficial similarities between RNAP traffic and vehicular 
traffic. We also derive an approximate analytical expression for this 
statistical analysis in a simplified special case and demonstrate its use 
by comparing with the corresponding data obtained from computer simulations.

Before presenting our quantitative model, we summarize a few essential 
steps in transcription. The RNAP locally unzips the two DNA strands 
creating a ``bubble'' whereby a single stranded DNA (ssDNA) template 
is exposed to it. Together with the DNA bubble and the growing RNA 
transcript, the RNAP forms a macromolecular complex called the 
``transcription elongation complex'' (TEC). The size of a single 
TEC is such that each incorporates $r$ successive nucleotides of the 
DNA template. During elongation, each mechano-chemical cycle of the 
RNAP consists of several steps. The major steps of this cycle involve 
the selection of the appropriate subunit for mRNA, as dictated by the 
DNA template and, then, its attachment to the growing mRNA transcript 
by a reaction that is catalyzed by the RNAP. Release of pyrophosphate 
($PP_{i}$), one of the products of this reaction is the rate-limiting 
step in each cycle. Thus, in each cycle, an RNAP steps forward by one 
nucleotide. The elongation process ends when the TEC encounters the 
corresponding ``termination sequence'' and the nascent mRNA is released 
by the RNAP.

Our model of transcription is shown schematically in fig.\ref{fig-model} 
where the essential components of each of the TECs are shown explicitly. 
The green and red squares at the start regions of the gene indicate the 
``ON'' and `OFF'' states of the gene, respectively. The rate constant 
(i.e., probability per unit time) of transition from the ``OFF'' state 
to the ``ON'' state is denoted by the symbol $\omega_{on}$ whereas that 
of the reverse transition is denoted by $\omega_{off}$. Initiation and 
termination of transcription are captured by the same prescription which 
have been used in our earlier work reported in ref.\cite{tripathi08}; 
the corresponding rate constants being $\omega_{\alpha}$ and 
$\omega_{\beta}$, respectively. 

In our model, the mechano-chemical cycle of individual RNAPs in the 
elongation stage and the nature of their steric interactions are  
identical to those used in ref.\cite{tripathi08}. For the sake of 
completeness, all the possible mechano-chemical transitions of an 
individual RNAP during the elongation stage are shown in 
fig.\ref{fig-mechchem1}(a).  Since pyrophosphate release is the rate 
limiting step, we assume that, at any given instant of time, a RNAP can 
exist in one of the two possible ``chemical'' states; no pyrophosphate 
($PP_{i}$) is bound to the RNAP in the state $1$ whereas the 
$PP_{i}$-bound state of the RNAP is labelled by the index $2$. 
The rate of $PP_{i}$ release is denoted by $\omega_{12}$ while the 
reverse reaction takes place at the rate $\omega_{21}$.  
The rate constants $\omega^{f}_{21}$, $\omega^{f}_{11}$ and 
$\omega^{f}_{22}$ correspond to polymerization of RNA whereas the  
rate constants $\omega^{b}_{12}$, $\omega^{b}_{11}$ and 
$\omega^{b}_{22}$ correspond to depolymerization of the RNA. 

For our numerical calculations, we have used the same set of rate constants
which we used in ref.\cite{tripathi08}; these are as follows:\\
\begin{eqnarray}
\omega^{f}_{21} &=& \omega^{f0}_{21} \cdot [NTP], ~{\rm with} ~\omega^{f0}_{21} = 10^{6} ~M^{-1} \cdot s^{-1} \nonumber \\ 
\omega^{f}_{11} &=& \omega^{f0}_{11} \cdot [NMP], ~{\rm with} ~\omega^{f0}_{11} = 46.6 ~M^{-1} \cdot s^{-1} \nonumber \\
\omega^{f}_{22} &=& \omega^{f0}_{22} \cdot [NMP], ~{\rm with} ~\omega^{f0}_{22} = 0.31 ~M^{-1} \cdot s^{-1} \nonumber \\
\omega_{21} &=& \omega^{0}_{21} \cdot [PP_{i}], ~{\rm with} ~\omega^{0}_{21} = 10^{6} ~M^{-1} \cdot s^{-1} \nonumber \\
\omega_{12} &=& 31.4 ~s^{-1} \nonumber \\
\omega^{b}_{12} &=& 0.21 ~s^{-1} \nonumber \\
\omega^{b}_{11} &=& 9.4 ~s^{-1} \nonumber \\
\omega^{b}_{22} &=& 0.063 ~s^{-1} \nonumber \\
\end{eqnarray}
where $[NTP]$, $[NMP]$ and $[PP_{i}]$ denote the concentrations of 
nucleoside triphosphate (NTP), nucleoside monophosphate (NMP) and 
pyrophosphate (PP$_{i}$), respectively. Moreover, for the figures  
in this letter, we have used $\omega_{\alpha}=5.0 s^{-1}$, 
$\omega_{\beta}=50   s^{-1}$, $\omega_{off}  =0.01 s^{-1}$, 
$\omega_{on} =0.001 s^{-1}$, and the concentrations $[NTP]=10^{-4} M$, 
$[PP_{i}] =10^{-6} M$, $[NMP]= 10^{-6} M$.

Note that, in 
spite of all the possible transitions shown in fig.\ref{fig-mechchem1}(a), 
the dominant pathway is the one shown in fig.\ref{fig-mechchem1}(b), where 
%the dominant pathway is the one shown in fig.\ref{fig-mechchem2}, where 
$\omega^{f}_{21}$ is proportional to the concentration of the available 
NTP subunits. 
All the quantitative predictions of our theory would remain equally 
valid if RNAP remains immobilzed while the template DNA passes through 
it in steps of one base pair, a scenario based on the concept of 
``transcriptional factory'' {\it in-vivo} \cite{cook99}. This alternative 
scenario would be mathematically related to the one used in this paper by  
just a coordinate transformation \cite{cozza06}- from the rest frame 
of the template DNA to that of the RNAP.

Mere visual examination of the time series of the transcriptional events 
(see Fig.\ref{fig-timeseries} for a typical one) establishes the 
occurrence of random bursts of transcriptional activities in our model 
\cite{golding06}.  
In order to sort these events into separate bursts, let us use a resolution   
$\Delta t$. Members of the same ``burst'' are separated from the immediate 
preceeding and suceeding transcriptional events by time gaps smaller than 
$\Delta t$ while the time gap between any pair of successive bursts is at 
least $\Delta t$ (or, longer). Our choice of $\Delta t = 2.5$ min. is 
motivated by the corresponding choice in typical laboratory experiments. 
We have also analyzed the same data using $\Delta t = 0.5$ min to test 
whether the conclusions drawn from our sorting procedure are, indeed, robust.

The number of transcriptional events in a burst is a measure of its 
{\it size}. The probability of the occurrence of a burst of size $n$ 
is given by 
\begin{equation}
P(n) = P_{on} ~p_{tr}^{n} ~P_{off},
\label{eq-expo1}
\end{equation}
where $P_{on}$ and $P_{off}$ are the probabilities of the gene switching 
ON and OFF, respectively, while $p_{tr}$ is the probability 
that a transcriptional event is completed by a RNAP. We can recast 
equation (\ref{eq-expo1}) into the exponential form 
\begin{equation}
P(n) = P_{on}~ P_{off}~ exp(-n/b),
\end{equation}
where $(1/b) = - ln~ p_{tr}$. Obviously, $b$ is the average size of a 
transcriptional burst; the higher is the magnitude of $p_{tr}$ the 
larger is the average size of the bursts. 

Our model goes beyond most of the earlier models of noise in transcription 
of a single gene because our model can predict the explicit dependence of 
$P_{on}$, $P_{off}$ and $p_{tr}$ on the rates of the steps of the 
mechano-chemical cycles of individual RNAPs as well as on their interactions. 
Suppose, $\omega_{eff}$ is the effective rate constant associated with 
the process of forward movement of the RNAP by one site (i.e., one 
nucleotide). Obviously, considering only the dominant pathway shown in 
fig.(\ref{fig-mechchem1}(b)) 
$\frac{1}{\omega_{eff}} ~=\frac{1}{~\omega_{12}} + \frac{1}{~\omega_{21}^f}$ 
and, hence,
$\omega_{eff} ~=\frac{~\omega_{12}~\omega_{21}^f}{~\omega_{12}~+~\omega_{21}^f}$.
An RNAP can attach to DNA strand only after the preceeding RNAP vacates 
the initial $r$ sites on the lattice. The rate at which a RNAP moves by 
$r$ sites is $k_{eff}=\frac{\omega_{eff}}{r}$.
Since $k_{eff}\ll\omega_{\alpha}$, the rate limting step in the process 
of transcription will be the initiation which will be determined essentially 
by $k_{eff}$. Hence,
$P_{tr} \propto exp\left( \frac{-1}{k_{eff}~\langle T_{on}\rangle}\right)$ 
where $\langle T_{on}\rangle~=~\frac{1}{\omega_{off}}$. Thus,
\begin{equation}
P_{tr} \propto exp\left( \frac{-\omega_{off}}{k_{eff}}\right) 
\label{eq-ptr}
\end{equation}
Moreover, 
\begin{equation}
P_{on}=\frac{\omega_{on}}{\omega_{on}+\omega_{off}}~ {\rm and}~ P_{off}=\frac{\omega_{off}}{\omega_{on}+\omega_{off}}
\label{eq-ponoff}
\end{equation} 
Finally, after normalization, the discrete distribution of the burst 
sizes is given by 
\begin{equation}
P(n) = \left( 1- exp\left( \frac{-\omega_{off}}{k_{eff}}\right) \right) \exp\left( \frac{-n~ \omega_{off}}{k_{eff}}\right).
\label{eq-sizeburst} 
\end{equation}

A typical distribution of the sizes of the bursts, obtained from 
computer simulations of our model, is plotted in Fig.\ref{fig-sizeburst} 
using two different values of $\Delta t$. These data are in excellent 
agreement with the theoretically predicted distribution 
(\ref{eq-sizeburst}); this exponential distribution is also consistent 
with the corresponding experimental observations \cite{golding05,chubb06}. 
Moreover, the data plotted in the inset of Fig.\ref{fig-sizeburst} 
also fit an exponential distribution thereby establishing that our 
conclusion is robust and independent of the actual magnitude of 
$\Delta t$ as long it remains within a reasonable range.

The duration of a burst is measured by the time interval between the 
first and the last transcriptional events which are members of the 
same burst. It is straightforward to see that the normalized 
distribution $P(t_{dur})$ of the burst durations $t_{dur}$ is given by 
\begin{equation}
P(t_{dur}) = \omega_{on}~exp\left( - \omega_{on}~t_{dur} \right)
\label{eq-durburst} 
\end{equation}
The theoretically predicted exponential distribution (\ref{eq-durburst}) 
is in excellent quantitative agreement with the  
corresponding numerical data obtained by direct computer simulations 
(see Fig.\ref{fig-durburst}). The experimental data reported by 
Chubb et al.\cite{chubb06} are also plotted in the inset of 
Fig.\ref{fig-durburst}. The nature of the distribution (namely, the 
exponential form) established by out theory and simulation is 
consistent with that observed in the experiments. The quantitative 
difference between the results predicted by our model and those 
obtained from experiments arises from the fact that the rate 
constants for the system used in the experiments are not necessarily 
identical to those used in plotting the results of our theory and 
simulation.

The time interval between a two successive bursts is the time gap 
between the last event of the earlier burst and the first event of 
the later burst. The normalized distribution $P(t_{int})$ of the 
intervals $t_{int}$ between successive bursts is given by 
\begin{equation}
P(t_{int}) = \omega_{off}~exp\left( - \omega_{off}~t_{int} \right)
\label{eq-intburst} 
\end{equation}
The quantitative agreement between this theoretical prediction and 
the corresponding simulation data (see 
Fig. \ref{fig-intburst}) is also consistent with the form of 
the distribution indicated by the experimental data reported by 
Chubb et al. \cite{chubb06}. However, because of the large 
scatters in the experimental data, no quantitative comparison 
between our theoretical predictions and experimental observations 
could be made.

Drawing an analogy to vehicular traffic \cite{css}, we define the {\it 
time headway} to be the time gap between the departures of the successive 
RNAPs from the termination site. Thus, according to this definition, 
the time-headway is the time gap between the completion of the 
synthesis of successive RNA molecules.  A typical distribution of the 
time headways is plotted in fig.\ref{fig-thrnap}. In the same figure 
we have also plotted the time headway distribution for a hypothetical 
scenario (which was considered in ref.\cite{tripathi08}) where the 
gene always remains ON. The best fit to both these curves are gamma 
functions (with slightly different parameters). A comparison between 
these two curves shows that the switching ON and OFF of the genes 
leads to a weak broadeing of the distribution; the longer tail caused 
by the gap between the successive bursts is shown separately in the 
inset of fig.\ref{fig-thrnap}.

We have been able to obtain an analytical estimate of the TH distribution 
only in a special limiting case exploiting the formal analogy with the 
models of vehicular traffic \cite{css}. 
Approximating the mechano-chemical cycle of each RNAP 
during the elongation stage by the pathway shown in fig.\ref{fig-mechchem1}(b), 
we can represent each RNAP (or, more precisely, each TEC) by a rigid 
rod, of length $r$, which can hop from one nucleotide to the next on the 
template DNA with an {\it effective} hopping probability $q$ per time 
step. Thus, 
\begin{equation} 
q \simeq \omega_{12} ~dt,  
\label{eq-qeff} 
\end{equation}
where $dt$ is the duration of each discretized time step. In 
this limit our original model reduces to the totally asymmetric simple 
exclusion process (TASEP) for hard rods of length $r$ \cite{schuetz}, 
provided the gene always remains in the ON state. In the special case 
$r = 1$ the rods reduce to particles and the corresponding exact TH 
distribution for TASEP (with parallel updating) is given by  
\cite{chow98a,chow98b} 
%\begin{widetext}
\begin{eqnarray}
{\cal P}_{\tau} &=&
\left[\frac{qy}{\rho-y}\right] \{1-(qy/\rho)\}^{t-1} \nonumber \\ 
&+& \left[\frac{qy}{(1-\rho)-y}\right] \{1-(qy/(1-\rho))\}^{t-1} \nonumber \\ 
&-& \left[\frac{qy}{\rho-y}+\frac{qy}{(1-\rho)-y}\right] p^{t-1} -
q^2(t-1)p^{t-2}, \nonumber \\
\label{eq-thtasep}
\end{eqnarray}
%\end{widetext}
where
\begin{equation}
y = \frac{1}{2q}\left( 1 - \sqrt{1 - 4 q \rho (1-\rho)}\right).
\end{equation}
and $\rho$ is the number density of the particles.

In order to test the range of validity of the expression (\ref{eq-thtasep}) 
in the context of RNAP traffic, we have carried out computer simulations 
of our model for $r = 1$ under periodic boundary conditions keeping the 
gene always ON. In the first set of simulations, we have used the 
simplified mechano-chemical cycle shown in fig.\ref{fig-mechchem1}(b) whereas in 
the second set we retained all the mechano-chemical transitions allowed 
in fig.\ref{fig-mechchem1}(a). The expression (\ref{eq-thtasep}) is in excellent 
agreement with the simulation data for the unbranched mechano-chemical cycle 
shown in fig.\ref{fig-mechchem1}(b). Moreover, even when the branched pathways 
of fig.\ref{fig-mechchem1}(a) exist, the simulation data are in reasonably good 
agreement with (\ref{eq-thtasep}).

In this letter we have reported a model that is ideally suited to study 
the effects of the steps of the mechano-chemical cycle of individual 
RNAPs and their steric interactions on the transcriptional bursts which 
are caused primarily by the switching of the gene between ``ON'' and 
``OFF'' states. For the sake of simplicity, we have illustrated our 
approach with a minimal model of RNAP mechano-chemistry which assigns 
only two possible distinct chemical states to an RNAP at any given 
location. For a more (biologically) realistic description, this 
mechano-chemistry of the RNAPs can be easily replaced by a more 
appropriate one without changing the overall framework of our model. 

Suppose $T_{obs}$ is the total time interval of observation 
and data collection in each single-cell experiment on transcriptional 
noise. In ref.\cite{tripathi08}, our theoretical analysis was restricted 
to a temporal regime such that (I) $T_{obs} < T_{on}$, where $T_{on}$ is 
the average duration for which the gene remains ON, 
and (II) $T_{obs} \ll T_{cell}$, where $T_{cell}$ is the mean life time 
of the cell before its division into the two daughter cells. Under these 
restrictions, our model of transcription \cite{tripathi08} did not exhibit 
transcriptional bursts. In this letter we have shown that the same model 
can account for transcriptional bursts when we relax the constraint (I). 
We show that the statistical properties of noisy transcription in our 
model in the temporal regime $T_{on} \ll T_{obs} < T_{cell}$ are in 
excellent agreement with the corresponding experimental data. Moreover, 
drawing an analogy to vehicular traffic, we have reanalyzed the time series 
of the transcriptional events from a totally different perspective which 
does not require any sorting of the raw data into separate bursts. 

This work is supported (through DC) by a research grant from CSIR
(India). DC also acknowledges hospitality of CCMT of IISc. Bangalore 
during the preparation of this manuscript.

%%%%%%%%%%%%%%%%%%%%%%%%%%%%%%%%%%%%%%%%%%%%%%%%%%%%%%%%%%%%%%%%%%%%%%%%

%%%%%%%%%%%%%%%%%%%%%%%%%%%%%%%%%%%%%%%%%%%%%%%%%%%%%%%%%%%%%%%%%%%%%%%%
\end{document}